# Ultimate Thin Vertical p-n Junction Composed of 2D Layered Molybdenum Disulfide

Hua-Min Li[+,*], Daeyeong Lee[#,*], Deshun Qu[#], Xiaochi Liu[#],

Jungjin Ryu[#], Alan Seabaugh[+], and Won Jong Yoo[#]

[#] SKKU Advanced Institute of Nano Technology (SAINT), Samsung-SKKU Graphene Center (SSGC), Department of Nano Science and Technology, Sungkyunkwan University (SKKU), Suwon, 440-746, Korea

[+] Center for Low Energy Systems Technology (LEAST), Department of Electrical Engineering, University of Notre Dame, Indiana, 46556, USA

[*] These authors contributed equally to this work.

Correspondence and requests for materials should be addressed to W. J. Y. (email: yoowj@skku.edu).

**Abstract**

Semiconducting 2D crystals are currently receiving significant attention due to their great potential to be an ultra-thin body for efficient electrostatic modulation which enables to overcome the limitations of silicon technology. Here we report that, as a key building block for 2D semiconductor devices, vertical p-n junctions are fabricated in ultrathin $MoS_2$ by introducing $AuCl_3$ and benzyl viologen dopants. Unlike usual unipolar $MoS_2$, the $MoS_2$ p-n junctions show (i) ambipolar carrier transport, (ii) current rectification via modulation of potential barrier in films thicker than 8 nm, and (iii) reversed current rectification via tunneling in films thinner than 8 nm. The ultimate thinness of the vertical p-n homogeneous junctions in $MoS_2$ is experimentally found to be 3 nm, and the chemical doping depth is found to be 1.5 nm. The ultrathin $MoS_2$ p-n junctions present a significant potential of the 2D crystals for flexible, transparent, high-efficiency electronic and optoelectronic applications.

**Index terms: molybdenum disulfide, p-n junction, tunneling, photoresponse**

Since the rediscovery of stable monolayer graphite or graphene, two-dimensional (2D) layered materials or van der Waals materials have led to remarkable interest in the physics and applications of the materials[1–3]. Graphene provides a variety of fascinating properties, including an ultrahigh carrier mobility, large mechanical strength, a linear dispersion relation, long-range ballistic transport, quantum hall effects at room temperature, and tunable optical absorption properties[4–6]. Beyond graphene, other 2D materials provide a rich variety of more flexible electronic properties, including wide band gap insulators, such as hexagonal boron nitride (*h*-BN)[7], semiconductors, and even superconductors, as may be observed in black phosphorus[8,9] or transition metal dichalcogenides (TMDCs)[10–12].

Unlike graphene, which cannot provide low current-off or saturated current-on states due to its zero band gap, the semiconducting TMDCs, such as n-type molybdenum disulfide ($MoS_2$), possess sizable band gaps in the range of 1–2 eV with sub-nanometer thickness, and provide high on/off ratios as well as more efficient control over switching[10–12]. $MoS_2$ has an indirect band gap of 1.3 eV in bulk structures but a direct band gap of 1.8 eV in monolayer form. The tunable electronic properties of $MoS_2$ enable electron tunneling and negative differential resistance (NDR) for use in low-power electronics[13,14]. The material is not susceptible to short-channel effects and this could be helpful in breaking through the scaling limits for transistor miniaturization[15–17]. Theoretical simulations indicate that a $MoS_2$ field effect transistor (FET) could operate in the ballistic regime to yield excellent device performances, including an on/off ratio of $10^{10}$ and a subthreshold swing (SS) of approximately 60 mV/dec[18]. $MoS_2$ and its hybrid heterostructures formed with other 2D materials have demonstrated significant potential for use in flexible, transparent, low-power electronics and optoelectronics, such as tunneling transistors[13], memories[19–21], photodetectors[22–25], electroluminescent devices[26], light-emitting devices[27], and integrated circuits[28,29]. Although the carrier mobility of $MoS_2$ is relatively low, it can be improved significantly by functionalizing the substrate[30], passivating the surface[31], applying high-$k$ dielectric engineering[32,33], or forming inversion channels[34].

Chemical doping has been shown to offer an effective approach to doping in electronic low-dimensional material applications including carbon nanotubes (CNTs), mono- or few-layer graphene[35]. Chemically doped TMDC materials and the applications of these materials, however, have not been extensively studied[36,37]. In this work, we successfully fabricated unipolar p-type doped $MoS_2$ (p-$MoS_2$), n-type doped $MoS_2$ (n-$MoS_2$), and pristine $MoS_2$ (pristine-$MoS_2$) FETs using chemical doping of gold chloride ($AuCl_3$) and benzyl viologen (BV).

We investigate the thickness-dependent electrical behavior of a vertical p-n homogeneous junction composed of $MoS_2$. The few-layer $MoS_2$ p-n junctions show ambipolar carrier transport. The potential barrier in a $MoS_2$ p-n junction can be effectively modulated in films with a thickness exceeding 8 nm as they are in conventional semiconductor p-n diodes, giving rise to current rectification in which carrier transport is permitted under a forward bias; however, films with a thickness of less than 8 nm, a "reversed" current rectification is clearly observed in which a tunneling-dominated current through the ultrathin potential barrier is favored under a reverse bias. The ultimate thickness and scaling limits of the vertical $MoS_2$ p-n junctions are experimentally determined to be 3 nm (4 layers). The chemical doping depth in the direction perpendicular to the layers is found to be 1.5 nm (2 layers). Reducing the film thickness below 3 nm, for example, in monolayer $MoS_2$, compromises the p- and n-type doping, and one type of doping eventually overwhelms other types throughout the entire flake. The small film thickness, on the order of one nanometer, renders the ultrathin vertical p-n junction of $MoS_2$ potentially useful in flexible, transparent, high-efficiency electronic and optoelectronic applications, such as phototransistors and solar cells.

**Results**

The effects of the chemical doping on carrier transport and device performance were investigated by fabricating and comparing the performances of p-$MoS_2$, n-$MoS_2$, and pristine-$MoS_2$ FETs (see the Supplementary Figs. S1 and S2). The excellent doping results made the fabrication of an ultrathin vertical $MoS_2$ p-n homogeneous junction possible. Figure 1(a)–1(d) showed the fabrication details of a vertical $MoS_2$ p-n junction. A few-layer $MoS_2$ flake with a thickness of 11 nm was obtained by mechanical exfoliation and was used as the channel in a back-gate FET. The bottom surface was doped to form an n-type semiconductor by introducing BV, and the top surface was doped to form a p-type semiconductor by introducing $AuCl_3$. A Cr/Pd (5 nm / 50 nm) top electrode and a Cr/Pd/Cr (5 nm / 50 nm / 5 nm) bottom electrode were contacted with the top and bottom surfaces of the $MoS_2$ flake, respectively, to provide symmetric metal contacts. Both optical microscopy and atomic force microscopy (AFM) images clearly revealed that the stacking structure was bottom electrode / $MoS_2$ / top electrode, as shown in Fig. 1(e)–1(g). The drain-to-source current ($I_D$) was characterized as a function of the drain and gate voltages ($V_D$ and $V_G$) using a semiconductor parameter analyzer. A monochromator (655 nm, 15 mW) and a standard solar simulator (AM1.5 spectrum) were combined with electrical measurements to test the photoresponse.

Compared to the unipolar $MoS_2$ films, such as p-$MoS_2$, n-$MoS_2$, and pristine-$MoS_2$, the $MoS_2$ p-n homogeneous junction provided several advantages. First, it provided a clear rectifying effect on carrier transport. The output characteristics revealed a current rectification ratio, defined as the ratio of the forward current to the reverse current, of approximately 100, and the theoretical fits suggested an ideality factor (*n*) of 1.6, as shown in Fig. 2(a). The p-n junction properties varied depending on the applied $V_D$, as illustrated by the energy band diagrams shown in Fig. 3(a)–3(d). At equilibrium, a potential barrier was established within the channel that prevented electron and hole transport from the source to drain. As with conventional semiconductor p-n diodes, the barrier height could be increased by applying a reverse bias ($V_D$ < 0 V), or it could be reduced by applying a forward bias ($V_D$ > 0 V), giving rise to a rectifying effect on carrier transport.

Secondly, unlike the p-MoS$_2$, n-MoS$_2$, and pristine-MoS$_2$ FETs, which showed unipolar carrier transport, ambipolar carrier transport with a hysteresis window of 60 V was observed in the p-n MoS$_2$ FET, as shown in Fig. 2(b). Electron and hole transport were attributed to the presence of n-MoS$_2$ at the bottom surface and p-MoS$_2$ at the top surface, respectively, as shown in Fig. 3(e) and 3(f). Under a positive $V_G$, the majority carriers were generated via accumulation, which were electrons generated at the bottom (n-MoS$_2$) and holes generated at the top (p-MoS$_2$) of MoS$_2$ p-n junction. By contrast, the minority carriers were generated via inversion under a negative $V_G$, which were holes generated at the bottom (n-MoS$_2$) and electrons generated at the top (p-MoS$_2$). In both cases, the current flow at a positive $V_D$ was contributed by electrons and holes with (i) lateral in-plane transport along the n-MoS$_2$ or p-MoS$_2$ layers, and (ii) vertical inter-layer tunneling. It should be noted that electron transport, with a maximum current of the order of 10 nA, was dominant over the hole transport, with a maximum current of the order of 0.1 nA. Because the charge carrier density was controlled by capacitive coupling to the back-gate, the modulation of electron transport in n-MoS$_2$ by the gate, which was close to the dielectric layer, was more effective. By contrast, hole transport was not effectively modulated by the gate in the p-MoS$_2$ due to the additional capacitance of the pristine-MoS$_2$ ($C_i$). Assuming that the capacitance for electron transport in n-MoS$_2$ ($C_n$) was equal to the oxide capacitance ($C_{ox}$), as expressed by $C_n = C_{ox} = \varepsilon_{ox}/t$, the capacitance for hole transport in p-MoS$_2$ ($C_p$) could be approximated as $C_p = (C_{ox}^{-1} + C_i^{-1})^{-1}$. Therefore, $C_p$ was smaller than $C_n$, which reduced the coupling between the hole carriers and the gate. Here, $\varepsilon_{ox}$ is the oxide permittivity, and $t$ is the oxide thickness. In addition to the electron current and hole current, a small current between those two was observed near zero gate bias (see Fig. 2(a) and 2(b)). This current was introduced by the pristine MoS$_2$. Since the chemical doping depth was only 1.5 nm (as discussed below) for both n-type and p-type doping, the MoS$_2$ moieties in the middle of a few-layer structure can be remained as pristine (n-type), and can form a p$^+$-n-n$^+$ multi-junction along the vertical direction. The current contribution from the middle pristine MoS$_2$ moieties was relatively small due to its low carrier density compared to those of the chemically doped MoS$_2$ moieties. The current map collected under dark conditions as a function of $V_D$ and $V_G$ indicated that electron transport proceeded at positive $V_G$ and hole transport at negative $V_G$, as shown in Fig. 2(c). Carrier multiplication and avalanche effects were clearly observed under a reverse bias ($V_D < 0$ V). The mapping of the corresponding photocurrent (PC), defined as the difference between the values of $I_D$ under dark or illuminated conditions, revealed two peaks under a positive $V_D$, as shown in Fig. 2(d). The positions and magnitudes of the PC peaks indicated electron and hole transport and reflected the presence of gate-controlled metal–semiconductor barrier modulation[38,39].

Thirdly, the MoS$_2$ p-n homogeneous junction had the potential to be made ultrathin, transparent, and flexible, and its vertical junction structure gave rise to a relatively large junction area that was beneficial for optoelectronic applications. For example, the strong PC generation at positive $V_G$ and $V_D$ suggested that the p-n MoS$_2$ FET could be used as a phototransistor for light detection, as shown in Fig. 4(a) and 4(b). Under a forward bias applied at $V_G = 60$ V and $V_D = 1$ V, the magnitude of $I_D$ under illumination ($I_{D,light}$) in the p-n junction was about two orders of magnitude larger than the magnitude of $I_D$ under dark conditions ($I_{D,dark}$). The time-resolved characteristics revealed a reliable photoresponse with a stabilized PC ON/OFF ratio of ~100. Moreover, the vertical MoS$_2$ p-n homogeneous junction was demonstrated to be useful in photovoltaic applications, as shown in Fig. 4(c) and 4(d). Under illumination with a standard solar simulator, the MoS$_2$ p-n junction functioned as a solar cell when the gate was grounded, and its energy conversion performance, including its efficiency ($\eta$), fill factor (*FF*), and photoresponsivity (*R*) were estimated to be 0.4%, 0.22, and 30 mA/W, respectively. Considering that the junction was only 11 nm thick, the chemically doped MoS$_2$ p-n junction could potentially be quite useful in future flexible, transparent, and high-efficiency optoelectronics if the device parameters, including the layer thickness, electrode layout, doping agent, and concentration, were optimized.

The vertical MoS$_2$ p-n homogeneous junction in this work showed its own natural advantages, compared to other solar energy harvesting devices based on MoS$_2$ p-n junction and MoS$_2$ hybrid systems, including lateral MoS$_2$ p-n junction[37], MoS$_2$-Au[40], MoS$_2$-graphene[41], MoS$_2$-WS$_2$[41], MoS$_2$-WSe$_2$[42], and MoS$_2$-Si[43] systems (see the Supplementary Table S1). For example, in contrast to the lateral MoS$_2$ p-n junction, the vertical p-n junction can provide a much larger planar junction area (or active area). This was very important for optoelectronic applications since the larger active area would absorb more photons, generate more photo-excited charge carriers, and increase the conversion efficiency. Compared to the heterogeneous systems, the homogeneous junction can provide the maximized carrier transport efficiency. The photo-excited charge carriers could be very easily lost at the heterogeneous interface due to a variety of factors, including the mismatch of the geometric morphology and lattice structure, the presence of the dangling bonds, surface defects, chemical residuals, absorbed H$_2$O and O$_2$ molecules etc. Those factors could result in a high contact resistance at the interface and a low carrier transport efficiency through the interface in the heterogeneous systems. In contrast, the homogeneous junction can naturally exclude all those deleterious factors, minimize the carrier lost through the junction, and maximize the carrier transport efficiency.

**Discussion**

We characterized the electrical and optoelectronic performances of a vertical p-n homogeneous junction formed by chemically doping in few-layer MoS$_2$ films. It was straightforward and interesting to investigate the thickness limits of a vertical p-n junction. A thickness-dependent study was carried out by fabricating a series of MoS$_2$ p-n junctions from few-layer MoS$_2$ films (18, 7, or 4 layers) or from the monolayer structure, as shown in Fig. 5. The potential barrier was varied as the MoS$_2$ film thickness decreased (see Fig. 5(a)). The 18-layer MoS$_2$ p-n junction behaved as a conventional semiconductor diode, with current rectification properties that allowed carrier transport to proceed under a forward bias due to a reduction in the potential barrier under a positive $V_D$ (see Fig. 5(b)). By contrast, conventional diode behavior was not observed in the 7-, 4-, and 1-layer MoS$_2$ p-n junctions, in which the thickness of the p-n junction, *i.e.*, the width of the potential barrier, was reduced to several nanometers or even less than 1 nm, and a large tunneling current was observed at a negative $V_D$ (see Fig. 5(c)–(e)). Under a low reverse bias, field-induced band bending was not severe, and direct tunneling (DT) dominated carrier transport. The DT current ($I_{D,DT}$) depended linearly on the bias according to[44,45]

$$I_{D,DT} = \frac{A_{eff}\sqrt{m_0\phi_B}q^2V_D}{h^2 d}\exp\left[\frac{-4\pi\sqrt{m_0\phi_B}d}{h}\right], \tag{1}$$

where $A_{eff}$ is the effective contact area, $\phi_B$ is the barrier height, $m_0$ is the free electron mass, $q$ is the electronic charge, $h$ is Planck's constant, and $d$ is the thickness of the MoS$_2$ film (barrier width). By contrast, the tunneling distance for electron transport from the drain to the source was further reduced by field-induced band bending under a high reverse bias, and Fowler–Nordheim tunneling (FNT) became dominant. The FNT current ($I_{D,FNT}$) followed a nonlinear relation to the bias according to[44,45]

$$I_{D,FNT} = \frac{A_{eff} q^3 m_0 V_D^2}{8\pi h \phi_B d^2 m^*} \exp\left[\frac{-8\pi\sqrt{2m^*}\phi_B^{\frac{3}{2}}d}{3hqV_D}\right], \quad (2)$$

where $m^*$ ($0.45m_0$) is the effective electron mass of $MoS_2$[18]. Equation 2 could be further expressed in a linear relation as

$$\ln\left(\frac{I_{D,FNT}}{V_D^2}\right) = \ln\left(\frac{A_{eff} q^3 m_0}{8\pi h \phi_B d^2 m^*}\right) - \frac{8\pi\sqrt{2m^*}\phi_B^{\frac{3}{2}}d}{3hqV_D}. \quad (3)$$

According to Eq. 3, $\ln(I_D/V_D^2)$ versus $1/V_D$ could be plotted for each different $MoS_2$ film thickness (see Fig. 5(f)–(i)). The strong linear dependence under a high bias suggested that FNT was dominant, and the logarithmic growth at a low bias indicated that DT was dominant. The effective value of $\phi_B$ for FNT was estimated from the slope of the linear fits, which increased from 0.14 to 0.35 eV as the $MoS_2$ film thickness decreased from 11 to 0.7 nm. The transition voltage from DT to FNT ($V_{D,trans}$) also increased from –0.6 to –0.1 V (see Fig. 5(j)), suggesting that a smaller bias was needed to trigger FNT as the $MoS_2$ p-n junction thickness decreased. Moreover, the current rectification ratio as a function of the $MoS_2$ film thickness clearly indicated a threshold transition between conventional rectification (with a rectification ratio > 1) and "reversed" rectification (with a rectification ratio < 1) at approximately 8 nm (12 layers, see Fig. 5(k)). In other words, the tunneling effects became dominant in vertical $MoS_2$ p-n homogeneous junction as the film thickness dropped below 8 nm.

The strong in-plane bonding and weak van der Waals inter-planar interactions yielded a chemical doping depth in $MoS_2$ that differed from that seen in conventional semiconductors. As demonstrated previously, the $MoS_2$ p-n junction showed ambipolar carrier transport as a result of enhanced hole transport by $AuCl_3$ and enhanced electron transport by BV.

Ambipolar carrier transport may be used as a fingerprint of a p-n junction. As the $MoS_2$ film thickness was reduced from 18 to 4 layers, ambipolar carrier transport remained, indicating the appropriate formation of a p-n junction; however, in the monolayer $MoS_2$, only unipolar electron transport was observed, as shown in Fig. 6. This result may reflect the overlap and recombination of both p- and n-type doping in the monolayer $MoS_2$, which eventually results in a single dominant doping type (n-type doping in this work) throughout the entire monolayer film. In other words, a vertical p-n homogeneous junction could not be formed in the monolayer $MoS_2$. We experimentally measured the thickness limit for a vertical $MoS_2$ p-n junction to be 3 nm (4 layers). The chemical doping depth along the direction perpendicular to the layers was estimated to be 1.5 nm (2 layers) for both p- and n-type doping. In order to confirm the doping depth, a direct observation of doping profile in $MoS_2$ flakes was made by using secondary ion mass spectroscopy (SIMS) (see the Supplementary Fig. S3). The doping depth was found to be 2 nm for p-type doping (see the depth profile of Au which was originated from $AuCl_3$), and to be 1.5 nm for n-type doping (see the depth profile of C and H which were originated from BV). Those results were consistent with the value (1.5 nm) estimated from the electrical measurement.

This finding was further supported by fabricating another monolayer $MoS_2$ p-n junction device using the same doping process but with double top electrodes and double bottom electrodes. This device was designed to confirm the carrier transport type at the top and bottom surfaces, respectively (see the Supplementary Fig. S4). Output characteristics showed "reversed" current rectification in which a tunneling-dominated large current was observed at the reversed bias. Transfer characteristics showed unipolar electron transport over a wide $V_G$ range. Both the features were consistent with the electrical behavior of another monolayer $MoS_2$ p-n junction (see Fig. 5(e) and Fig. 6(d)), suggesting the good reproducibility and reliability of the vertical $MoS_2$ p-n junction in this work. The individual transfer characteristics on both the top and bottom surfaces showed electron-dominated carrier transport, suggesting the compromise of p-type doping and the overwhelming of n-type doping. This also agreed with our theory. To quantitatively analyze the metal–semiconductor contact condition, the metal–semiconductor barrier height ($\phi_{MS}$) was obtained by applying a temperature-dependent test. The maximum value of $\phi_{MS}$ obtained from both the top and bottom metal-semiconductor interfaces were about 40 meV at the positive $V_G$, which was in agreement with our previous discussion on the electrical behavior of a Schottky-like junction (see the Supplementary Fig. S2). Our work experimentally revealed the thickness limit of a vertical $MoS_2$ p-n homogeneous junction and established the scaling limit for use in further design and development.

In conclusion, both the unipolar $MoS_2$, such as the p-$MoS_2$ and n-$MoS_2$, as well as the ambipolar vertical $MoS_2$ p-n homogeneous junction were successfully fabricated by chemically doping $AuCl_3$ and BV. The thickness-dependent properties of the vertical $MoS_2$ p-n junction suggested that normal diode behavior occurred for a $MoS_2$ film thickness exceeding 8 nm, and tunneling-dominated "reversed" rectification occurred for a film thickness smaller than 8 nm. The ultimate thickness and scaling limits for the vertical $MoS_2$ p-n homogeneous junction were experimentally found to be 3 nm, and the chemical doping depth was found to be 1.5 nm. Given the small thickness, of the order of one nanometer, the vertical $MoS_2$ p-n homogeneous junctions potentially have significant utility in flexible, transparent, high-efficiency electronic and optoelectronic applications.

## Methods

**Device fabrication**

The fabrication of the p-MoS$_2$, n-MoS$_2$, and pristine-MoS$_2$ FETs began with mechanical exfoliation from bulk crystals. After transfer to a p-type Si substrate (1.0–10.0 Ωcm) coated with a 90 nm thick thermal oxide layer, the MoS$_2$ flakes were carefully selected by optical microscopy and AFM to have an approximate thickness of 10 nm for use in comparative studies. The p-MoS$_2$ or n-MoS$_2$ films were obtained by spin-coating a layer of AuCl$_3$ (20 mM) or BV (20 mM), respectively, followed by annealing on a hot plate at 100°C for 10 min. The pristine-MoS$_2$ sample reserved untreated as a reference sample. Metal Cr/Pd (5 nm / 50 nm) source and drain contact electrodes were patterned using standard electron beam lithography (EBL) and electron beam evaporation techniques.

The p-n MoS$_2$ FET was fabricated as shown in Fig. 1. Firstly, a MoS$_2$ flake was exfoliated from the bulk crystal onto a Si substrate, onto the surface of which had been spin-coated a water-soluble polyvinyl alcohol (PVA) layer and a hydrophobic polymethyl methacrylate (PMMA) film[7]. Then, onto the top surface of the MoS$_2$ flake was spin-coated a BV (20 mM) layer, and the assembly was annealed on a hot plate at 100°C for 10 min to form an n-MoS$_2$ surface (see Fig. 1(a)). Next, the Si substrate supporting the n-MoS$_2$ flake was floated on the surface of a deionized water bath. Once the PVA layer had completely dissolved, the PMMA film was left floating on top of the water and could be transferred to a glass slide, the surface of which was coated with a thick polydimethylsiloxane (PDMS) film. Then, the glass slide was clamped onto the arm of a micromanipulator mounted on an optical microscope. The MoS$_2$ flake was optically aligned with the n-MoS$_2$ surface downward and was precisely stacked on a bottom electrode that had been deposited in advance onto a target p-type Si substrate (1.0–10.0 Ωcm) coated with a 285 nm thick thermal oxide layer using standard EBL and electron beam evaporation techniques (see Fig. 1(b)). During the transfer process, the target substrate was heated to 135°C to drive off any water absorbed on the flake surface, as well as to promote adhesion between PMMA and the target substrate. After transfer, the PMMA and MoS$_2$ flake remained on the target substrate, and the PMMA layer was dissolved in acetone (see Fig. 1(c)). Next, onto the top surface of the MoS$_2$ flake was spin-coated a AuCl$_3$ (20 mM) layer. The structure was then annealed on a hot plate at 100°C for 10 min to form a p-MoS$_2$ surface. The top electrode was patterned using standard EBL and electron beam evaporation techniques (see Fig. 1(d)). The bottom electrode was composed of Cr/Pd/Cr (5 nm / 50 nm / 5 nm), and the top electrode was composed of Cr/Pd (5 nm / 50 nm) in order to provide symmetric metal contacts to p-n MoS$_2$ that were identical to the metal contacts used in the p-MoS$_2$ and n-MoS$_2$ devices, for comparison.

**Device measurements**

The electrical properties were characterized using a semiconductor parameter analyzer under vacuum conditions (10 mTorr) at room temperature. The source and drain contacts were equivalent in the unipolar p-MoS$_2$, n-MoS$_2$, and pristine-MoS$_2$ FETs. The bottom electrode in contact with the n-doped MoS$_2$ in the ambipolar MoS$_2$ p-n junction was set as the source and were grounded during all measurements. The top electrode in contact with the p-doped MoS$_2$ in MoS$_2$ p-n junction was set as the drain, and a drain bias was applied.

The optoelectronic properties were characterized using a monochromator (655 nm, 15 mW) in the phototransistor applications, and using a standard solar simulator (AM1.5 solar spectrum) in the solar cell application.

**Energy conversion performance**

In solar cell applications, the vertical MoS$_2$ p-n junction showed a short-circuit current ($I_{SC}$) of 5.1 nA, and an open-circuit voltage ($V_{OC}$) of 0.6 V. The current and voltage obtained at the maximum output power ($I_{max}$ and $V_{max}$) were 2.2 nA and 0.3 V, respectively. Given the vertical p-n junction area ($A$), which was estimated from the optical microscopy image to be ~170 μm$^2$, the short-circuit current density ($J_{SC}$) could be approximated according to $J_{SC} = I_{SC} / A = 3.0$ mA/cm$^2$, and the current density at the maximum output power ($J_{max}$) could be approximated according to $J_{max} = I_{max} / A = 1.3$ mA/cm$^2$. Assuming that the input power was equivalent to the solar spectrum ($P_{in}$) at 0.1 W/cm$^2$, the maximum output power ($P_{max}$) was estimated to be $P_{max} = J_{max}V_{max} = 0.4$ mW/cm$^2$, the energy conversion efficiency ($\eta$) was estimated to be $\eta = P_{max} / P_{in} = 0.4\%$, the fill factor ($FF$) was estimated to be $FF = P_{max} / (J_{SC}V_{OC}) = 0.22$, and the photoresponsivity ($R$) was estimated to be $R = J_{max} / P_{in} = 30$ mA/W.

**Acknowledgements**

This work is supported by the Basic Science Research Program through the National Research Foundation of Korea (NRF) (2009-0083540, 2013-015516), and by the Global Frontier R&D Program (2013-073298) on the Center for Hybrid Interface Materials (HIM), funded by the Ministry of Science, ICT & Future Planning.

**Author contributions**

H. M. L., A. S., and W. J. Y. conceived of the research project, supervised the experiment, and wrote the paper. H. M. L., D. L., and J. R. performed the device fabrication. H. M. L. and X. L. performed the electrical and optoelectronic characterization. H. M. L. and D. Q. performed the doping process and AFM analysis.

**Additional information**

**Competing financial interests:** The authors declare no competing financial interests.

# Figures

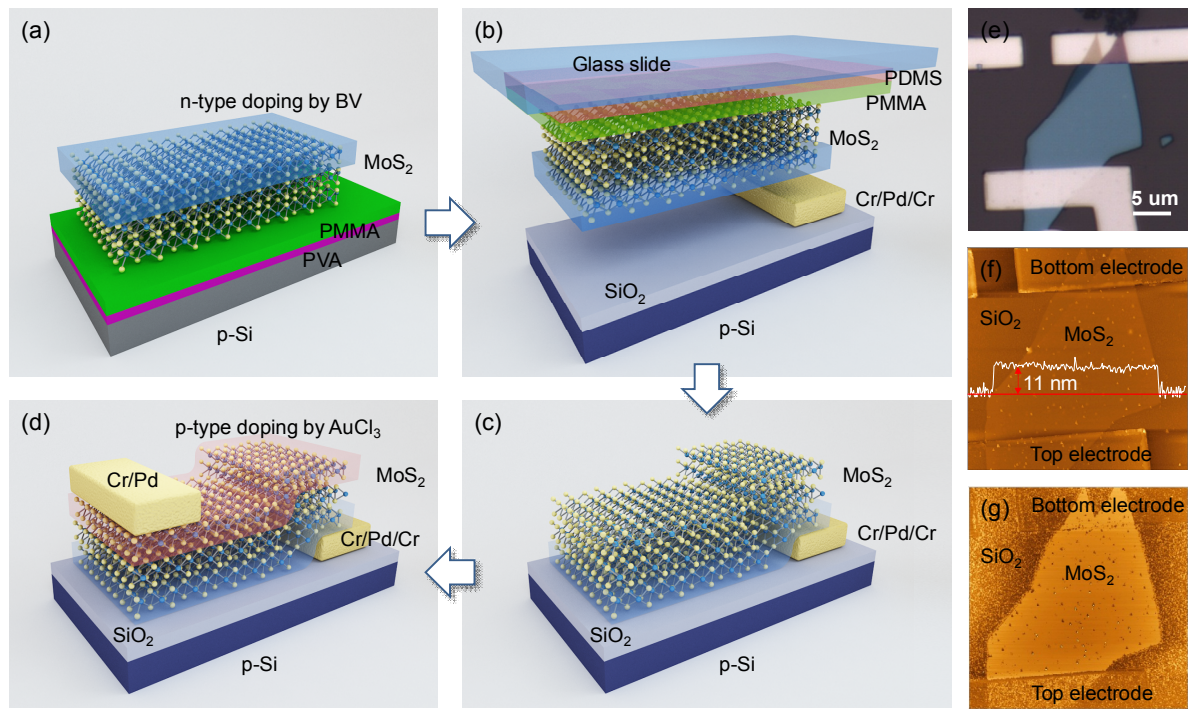

FIG. 1. **Fabrication of chemically doped vertical p-n homogeneous junction in a few-layer MoS$_2$ flake.** (a) A MoS$_2$ flake was transferred onto a PMMA/PVA/Si substrate, then BV-doped and annealed. (b) After dissolving the PVA layer in deionized water, the PMMA film supporting a MoS$_2$ flake was transferred to a PDMS/glass substrate. (c) The MoS$_2$ flake was stamped onto the SiO$_2$/Si substrate, and the n-doped surface was aligned with the Cr/Pd/Cr bottom electrode prepared in advance. (d) After AuCl$_3$ doping and annealing, the vertical p-n junction in the MoS$_2$ flake was formed, followed by the deposition of a Cr/Pd top electrode. (e, f, g) Optical microscopy image, AFM height image with a line scan profile, and AFM phase image of a vertical p-n homogeneous junction composed of a few-layer MoS$_2$ flake.

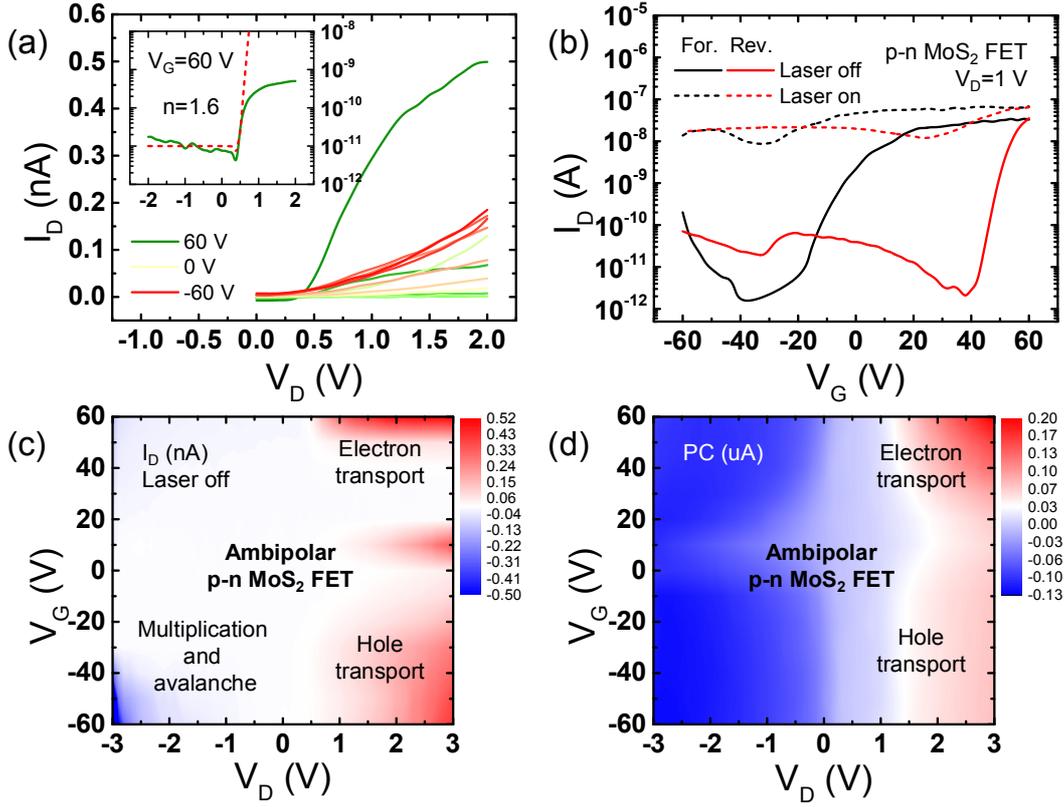

FIG. 2. **Electrical and optoelectronic properties of the p-n MoS$_2$ FET.** (a) Output characteristics at various $V_G$ levels between 60 and –60 V, along steps of 10 V. Inset: Output characteristics on the logarithmic scale in the current-on state. The ideality factor was estimated as 1.6. (b) The transfer characteristics and their photoresponses during both the forward and reverse sweeps. (c, d) Channel current mapping under dark conditions and the corresponding PC mapping as a function of various $V_D$ (from –3 to 3 V) and $V_G$ (from –60 to 60 V) levels illustrate the ambipolar carrier transport.

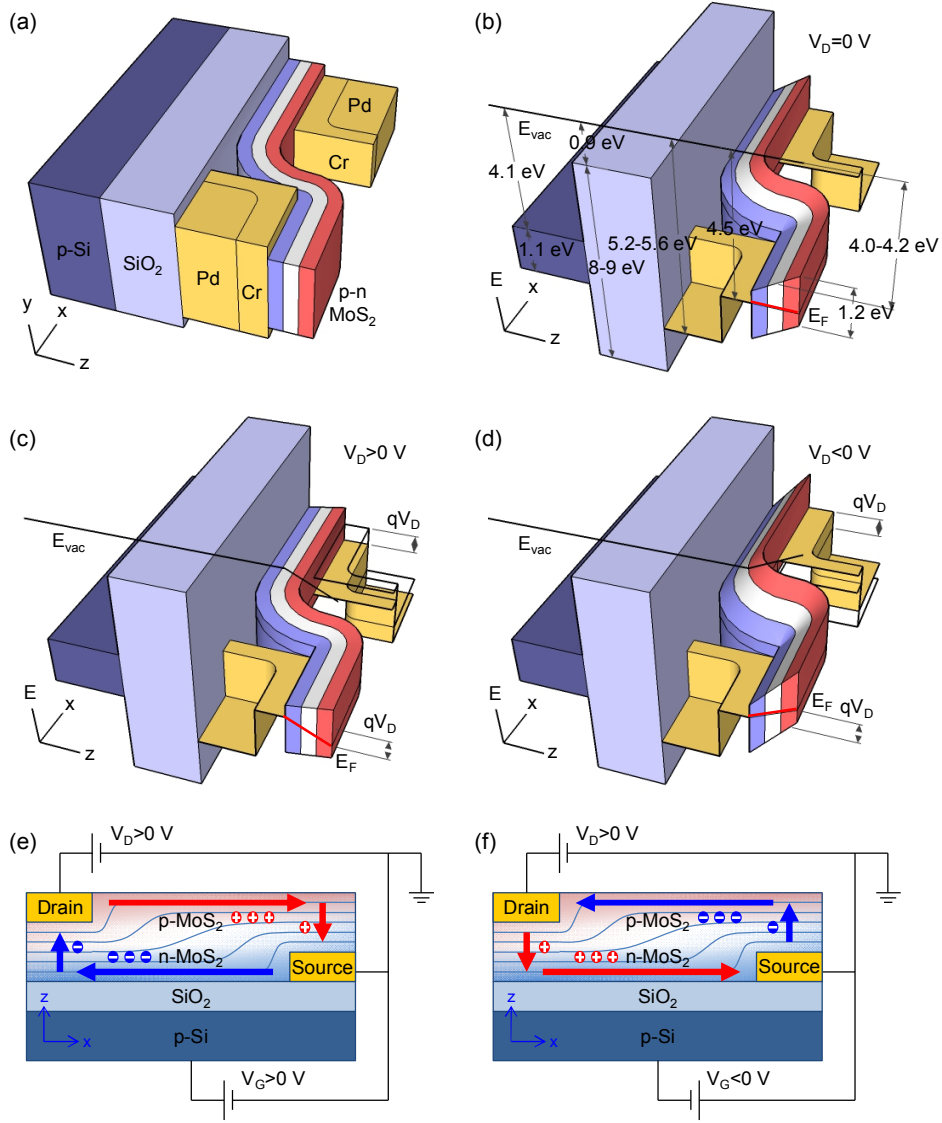

FIG. 3. **Effect of drain and gate biases on carrier transport in the p-n MoS$_2$ FET.** (a, b) The schematic diagram and the corresponding energy band diagrams versus the x-z plane under equilibrium condition. The black and red sold lines denote $E_{vac}$ along the z-axis and $E_F$ in the MoS$_2$ p-n junction, respectively. (c, d) The energy band diagrams illustrate a reduced potential barrier under a forward bias ($V_D > 0$ V), and an enlarged potential barrier under a reverse bias ($V_D < 0$ V). (e, f) The cross-section views illustrate the majority carrier transport at the accumulation ($V_G > 0$ V), and the minority carrier transport at the inversion ($V_G < 0$ V).

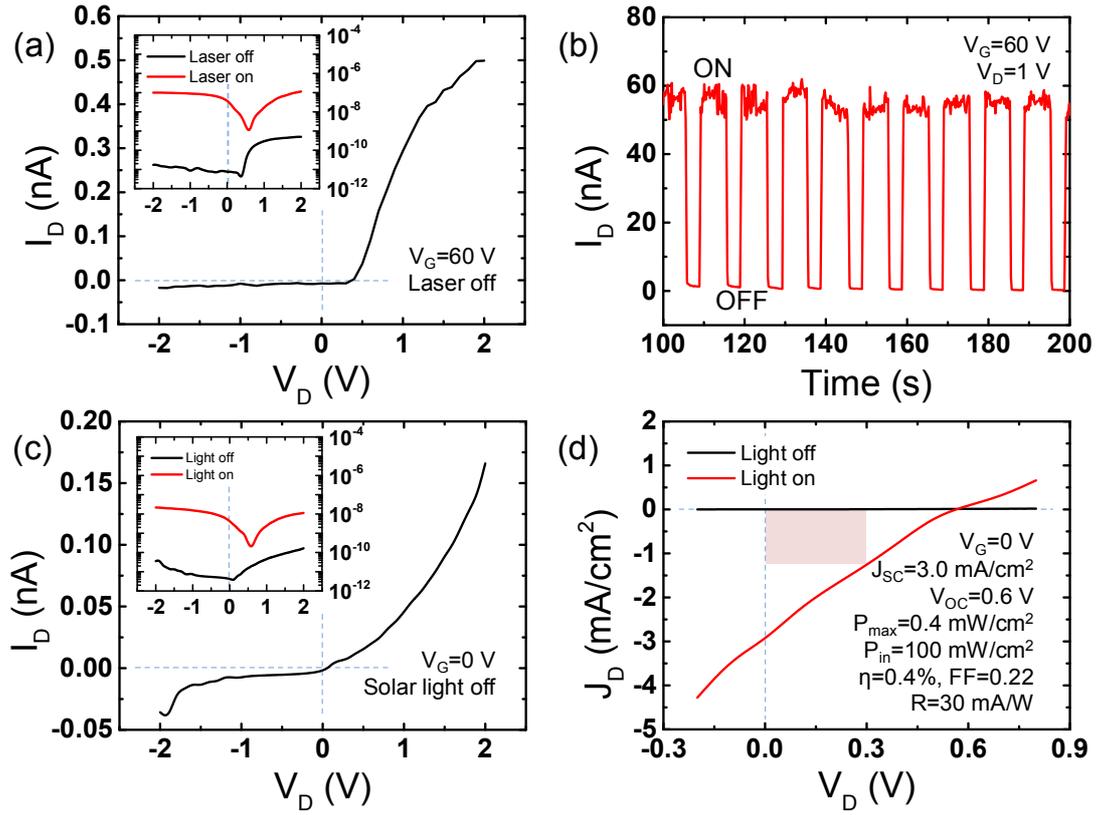

FIG. 4. **Application of vertical MoS$_2$ p-n junctions for use in optoelectronic applications.** (a, b) The p-n MoS$_2$ FET was used as a phototransistor for photodetection at $V_G$ = 60 V. The time-resolved photoresponse at $V_G$ = 60 V and $V_D$ = 1 V illustrates a PC ON/OFF ratio of ~100. (c, d) The MoS$_2$ p-n junction was used as a solar cell for light harvesting at $V_G$ = 0 V. The current density as a function of $V_D$ illustrates the energy conversion properties.

FIG. 5. **Thickness-dependent current rectification of vertical MoS$_2$ p-n junctions.** (a) Energy band diagrams of the devices prepared with vertical p-n junctions of various MoS$_2$ thicknesses. The MoS$_2$ band gap was equal to 1.2 eV for the few-layer structure

and 1.8 eV for the monolayer. (b, c, d, e) Output characteristics of the p-n MoS$_2$ FETs with layer numbers of 18, 7, 4, and 1. The red and blue backgrounds indicate the current-on and current-off states, respectively. (f, g, h, i) The corresponding Fowler–Nordheim plots of the vertical MoS$_2$ p-n junctions with layer numbers of 18, 7, 4, and 1. The red line denotes the linear fit to the FNT currents. (j) The barrier height and DT-FNT transition voltage as functions of the MoS$_2$ thickness and layer number. (k) Current rectification ratio as a function of the MoS$_2$ thickness and layer number at various $V_D$ (±3, ±2 and ±1 V) levels, indicating a transition between the conventional rectification and reversed rectification at ~8 nm (red dot circle).

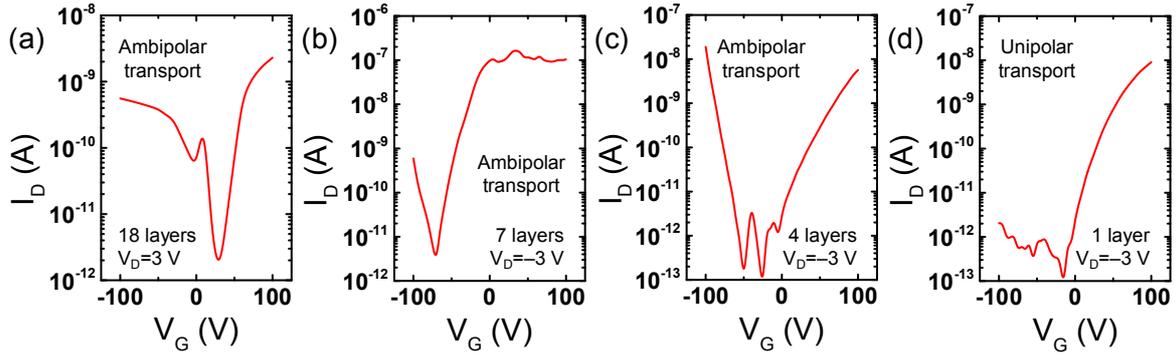

FIG. 6.  **Thickness-dependent carrier transport in MoS$_2$ p-n junctions.** Transfer characteristics of the vertical MoS$_2$ p-n junctions in the current-on state illustrate ambipolar transport for layer numbers of (a) 18, (b) 7, and (c) 4, but illustrate unipolar transport for (d) the monolayer.